\documentclass[preprint,proceedings]{rmaa}

\suppressfulladdresses 

\usepackage{paralist}

\usepackage{psfrag,color}

\newcommand{\HII}{H\,{\sc ii}}
\newcommand{\PAH}{{\sc PAH}}
\newcommand{\SIV}{[S\,{\sc iv}]}

\SetYear{2006}
\SetConfTitle{}

\title{Mid-Infrared T-ReCS Spectroscopy of Local LIRGs}

\author{
  T. D\'iaz-Santos,\altaffilmark{1} 
  A. Alonso-Herrero,\altaffilmark{1}
  L. Colina,\altaffilmark{1}
  C. Packham,\altaffilmark{2}
  J. T. Radomski,\altaffilmark{3}
  and C. M. Telesco\altaffilmark{2}
}

\altaffiltext{1}{Departamento  de 
Astrof\'isica Molecular e Infrarroja, IEM, CSIC, Madrid, Spain (tanio, aalonso, colina@damir.iem.csic.es).}

\altaffiltext{2}{Department of Astronomy, University of Florida, 
Gainesville, FL 32611, USA (packham, telesco@astro.ufl.edu).}
\altaffiltext{3}{Gemini Observatory, La Serena, Chile (jradomski@gemini.edu).}

\shortauthor{D\'iaz-Santos et al.}
\shorttitle{Mid-Infrared Spectroscopy of LIRGs}

\listofauthors{T. D\'iaz-Santos, A. Alonso-Herrero, L. Colina, C. Packham, J. T. Radomski, \& C. M. Telesco}
\indexauthor{D\'iaz-Santos, T.}
\indexauthor{Alonso-Herrero, A.}
\indexauthor{Colina, L.}
\indexauthor{Packham, C.}
\indexauthor{Radomski, J. T.}
\indexauthor{Telesco, C. M.}

\abstract{We present T-ReCS high spatial resolution $N$-band ($8-13\,\mu$m) 
spectroscopy of the central regions (a few kpc) of 3 local 
LIRGs. The nuclear spectra show deep 9.7\,$\mu$m silicate
  absorption feature and the high ionization \SIV$10.5\,\mu$m, 
emission line,  consistent with their optical classification as AGN. 
The two LIRGs with unresolved mid-IR emission do not show PAH emission
at $11.3\,\mu$m in their nuclear spectra. 
The spatially resolved mid-IR spectroscopy of NGC~5135 allows us to separate
out the spectra of the Seyfert nucleus, an \HII\ region, and the 
diffuse region between them on scales of less than $2.5\arcsec \sim
600\,$pc. The diffuse region spectrum is characterized by strong PAH emission
with almost no continuum, whereas the \HII\, region shows PAH emission with a
smaller equivalent width as well as [Ne\,{\sc ii}]$12.8\,\mu$m line. } 

\resumen{Se presenta espectroscopia de alta resoluci\'on espacial con T-ReCS
  en la banda $N$ ($8-13\,\mu$m) de las regiones centrales (unos pocos kpc) de
  tres galaxias   infrarrojas (IR) luminosas (LIRGs) locales. 
Los espectros nucleares presentan 
una profunda banda de absorci\'on de los silicatos a
  9.7\,$\mu$m y la l\'{\i}nea de emisi\'on de \SIV$10.5\,\mu$m en 
acuerdo con su clasificaci\'on como AGN. Las dos LIRGs con emisi\'on en el IR
  medio no resuelta no muestran emisi\'on PAH en $11.3\,\mu$m 
en sus espectros nucleares. Para NGC~5135  los datos de
  T-ReCS nos han permitido resolver los espectros del n\'ucleo, una regi\'on
  \HII\, y la zona difusa entre ambos, en escalas menores que $2.5\arcsec
  \sim  600\,$pc.  La zona difusa presenta emisi\'on PAH brillante con un
  continuo d\'ebil, mientras que la regi\'on H\,{\sc ii} presenta emisi\'on PAH
  con menor anchura equivalente y la l\'{\i}nea de [Ne\,{\sc
  ii}]$12.8\,\mu$m. }

\addkeyword{Galaxies: Infrared}
\addkeyword{Galaxies: Seyfert}
\addkeyword{Galaxies: Star Formation}
\addkeyword{H~II regions}

\begin{document}
\maketitle

\section{Introduction}
\label{sec:intro}

Luminous Infrared (IR) 
Galaxies (LIRGs, $10^{11} L_\odot \leq$L$_{\rm [8-1000\,\mu m]}
\leq\,10^{12} L_\odot$, 
Sanders \& Mirabel 1996) dominate the star formation (SF) rate density
(P\'erez-Gonz\'alez et al. 2005) at $z \sim 1$. 
The integrated properties of these high-$z$ LIRGs are similar to their local
universe analogs. 
Much of our knowledge of the mid-IR spectroscopic properties of local IR-bright galaxies
comes from {\it ISO} (e.g., Genzel et al. 1998) and early
results with IRS on {\it Spitzer} (Armus et al. 2004). However the
majority of the {\it ISO} works focused on samples of local IR-bright 
starburst galaxies (Verma et al. 2003), or included mainly ULIRGs (e.g.,
Genzel et 
al. 1998). There has been  a few ground-based high spatial resolution 
studies of  
LIRGs (e.g., Soifer et al. 2002) but again they tend
to study only the most luminous and famous examples in the LIRG class. 

In this paper we present the first results of a T-ReCS (Telesco et al. 1998)
spectroscopic study of three local LIRGs selected from the representative
sample of local LIRGs of Alonso-Herrero et al. (2006a). For this
sample we have already obtained Gemini-South/T-ReCS
mid-IR imaging data (Alonso-Herrero et al. 2006b) which allowed us
to initially select LIRGs  with bright compact nuclear emission so high S/N
spectra could be obtained. This tends to
include those LIRGs in our sample hosting an AGN, although as we shall see,
the superior spatial resolution achieved by T-ReCS allows us to separate the
AGN 
emission from the circumnuclear emission (see \S4).

\section{Observations}
\label{sec:obs}

We obtained T-ReCS $N$-band spectroscopy of the central 
(a few kpc) regions of NGC~5135, IC~4518W, and NGC~7130.
We used the low-resolution mode ($R\sim 100$) with a slit width of
$0\farcs72$. The pixel size of 0.089\arcsec  \, provides a 
slit of 21.6\arcsec \, in length. The on-source integration times were 30\,minutes per
target. The observing conditions were excellent making the T-ReCS observations
effectively diffraction limited (FWHM$\sim 0\farcs35$).
The 1D-spectrum extraction (see Figs.~1 and 2) for 
point sources was made $\lambda$-dependant, to account for the PSF variations, 
with an aperture of  
0\farcs36 at $\lambda = 10.37\,\mu$m.

\begin{figure}[!t]\centering
  \includegraphics[width=0.9\columnwidth]{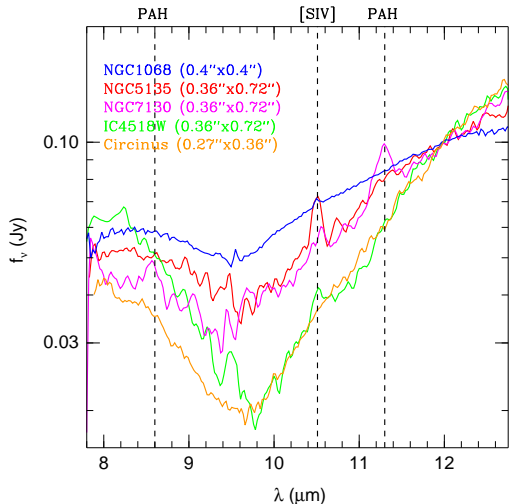}
\vspace{-0.5cm}
  \caption{T-ReCS rest-frame nuclear spectra of our 
three LIRGs: NGC~5135, IC~4518W, and
  NGC~7130, as well as two comparison Sy 2 galaxies:
  Circinus and   NGC~1068, shown in the order (top to bottom) 
given by the labels.  
The spectra have been normalized to 0.1~Jy at 12\,$\mu$m for a
  better 
  comparison of the depth of the 9.7\,$\mu$m silicate feature. The
  scaling factors of the spectra are: 1.23, 0.92, and 0.90 for NGC~5135,
  IC~4518W, and NGC~7130, and 0.008 and 0.07  for NGC~1068 and Circinus.} 
  \label{f:fig1}
\end{figure}


\section{Nuclear Emission}
\label{sec:seyf}

The 3 LIRGs  host an optically identified
AGN (NGC~5135 and IC~4518W are Seyfert (Sy) 2s, whereas the NGC~7130 has been
classified as Sy 2 or LINER).  
From our T-ReCS and  NICMOS 
imaging data we have detected unresolved nuclear emission in  NGC~5135
and IC~4518W, implying sizes of $< 40-51$\,pc, respectively, 
whereas the mid-IR nuclear emission of NGC~7130 appears
resolved with a size of $\sim 190\,$pc (see Alonso-Herrero et al. 2006b).
In addition we have detected in NGC~5135 and NGC~7130
bright \HII\, regions within the central $1-2\,$kpc.
The nuclear T-ReCS spectra of these LIRGs are presented in
Fig.~\ref{f:fig1}. The T-ReCS slit width covers
approximately the central 200\,pc. For comparison 
we include T-ReCS nuclear spectra of 2 bright nearby
Sy 2 galaxies for which the T-ReCS 
slit probes smaller physical regions ($\sim 30\,$pc for NGC~1068 Mason et
al. 2006 and  $\sim 6\,$pc for Circinus Roche et al. 2006). 

\begin{figure}[!t]
  \includegraphics[width=1.02\columnwidth]{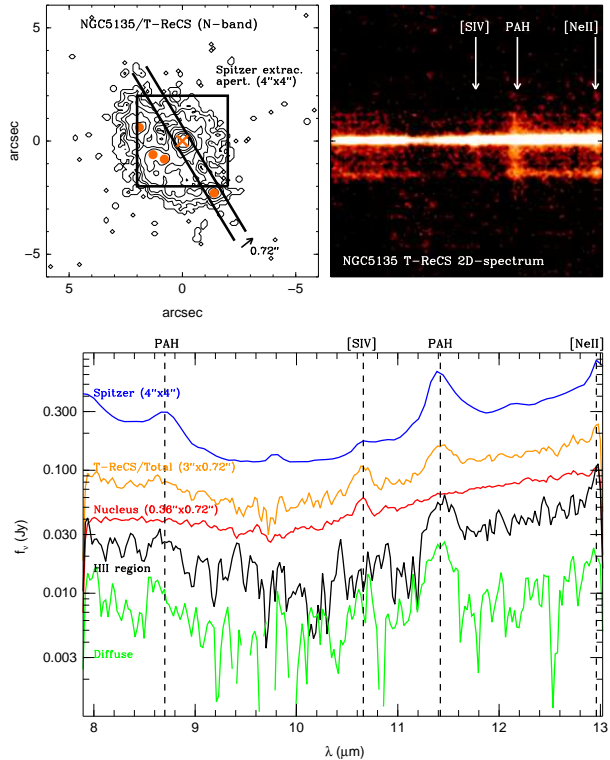}

\vspace{-0.5cm}
  \caption{{\it Upper left panel}: Contour plot at $10.4\,\mu$m of the 
central region of NGC~5135. The solid lines indicate the position and
orientation of 
the T-ReCS slit. The square indicates the approximate size of the extraction
  aperture of the {\it Spitzer}/IRS spectrum (see bottom panel). The cross
  marks the 
  Sy 2 nucleus, whereas the filled circles indicate the locations of
  luminous H\,{\sc ii} regions detected both in the mid-IR and Pa$\alpha$ (see
  Alonso-Herrero et al. 2006b for more details).  {\it Upper right panel}:
  2D T-ReCS spectrum of NGC5135. {\it  Lower panel}: Low-res IRS
spectrum extracted with the  smallest possible  aperture, and 
T-ReCS spectra for
the three regions discussed in the text: nucleus, H\,{\sc ii} region, and
diffuse region.} 
  \label{f:fig2}
\end{figure}

\begin{figure*}[!t]
\hspace{2cm}
  \includegraphics[width=0.8\columnwidth]{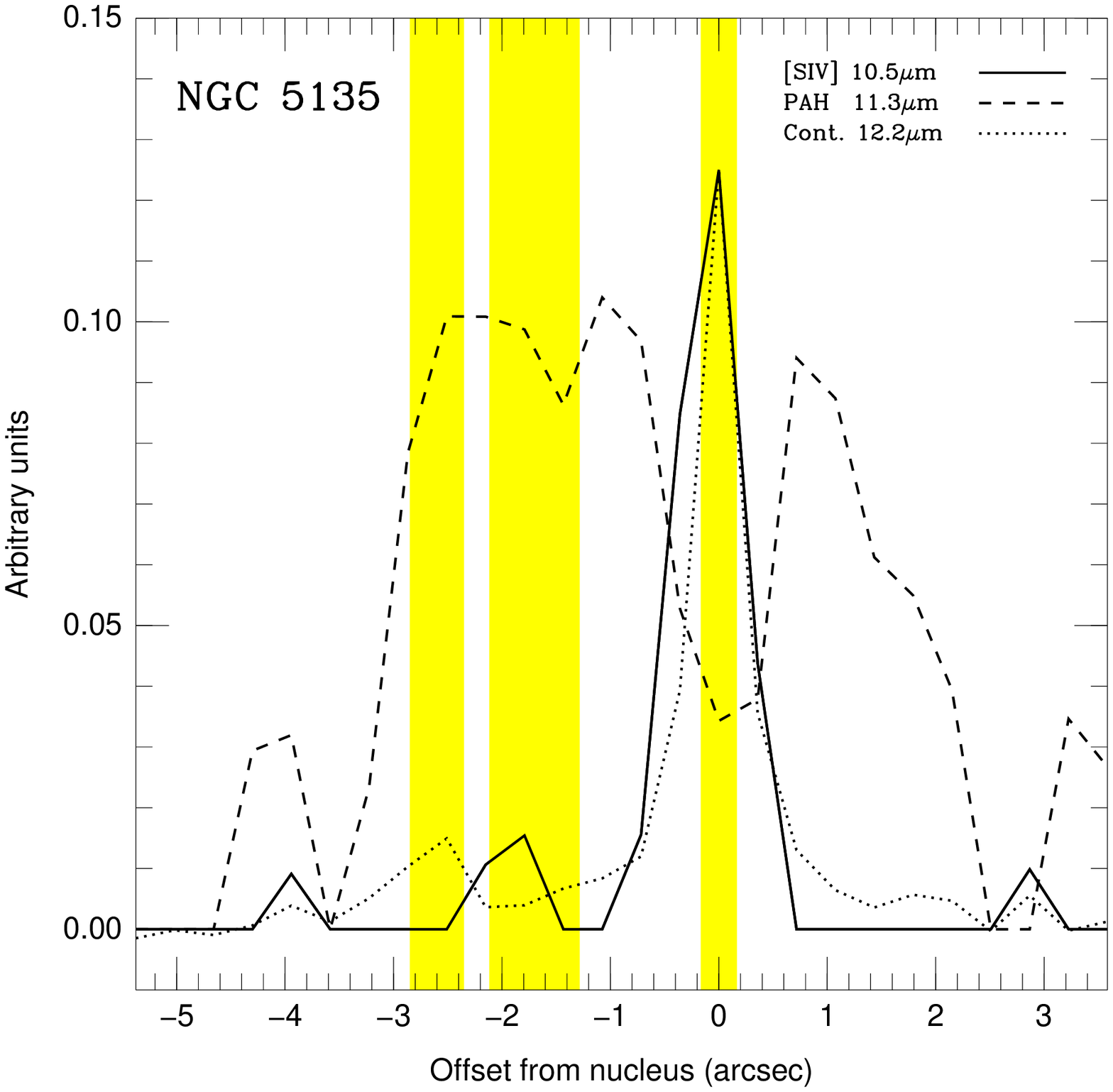}
  \includegraphics[width=0.8\columnwidth]{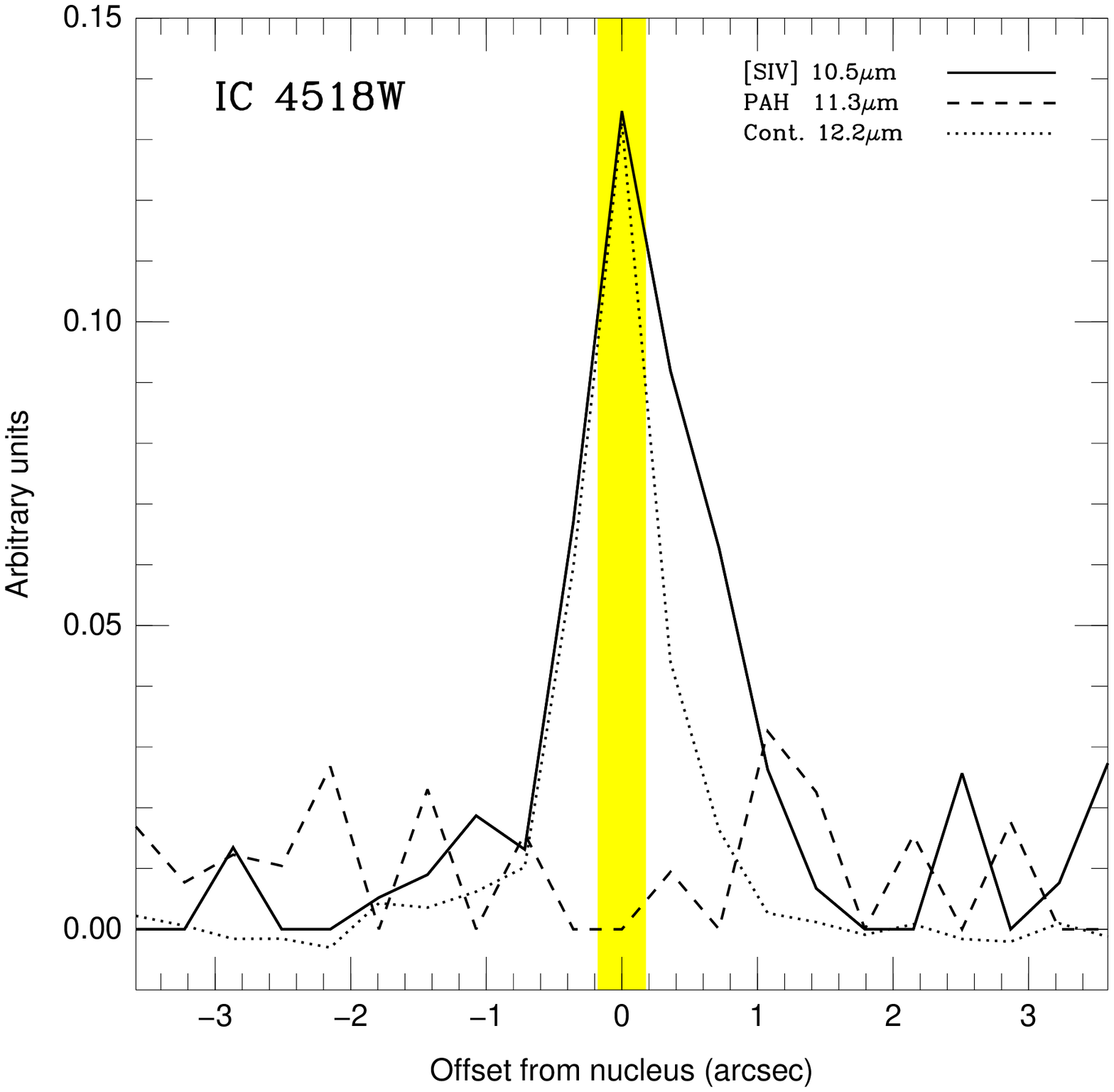}
\vspace{-0.5cm}
  \caption{T-ReCS 
spatial profiles of the \SIV \, line, the $11.3\,\mu$m
  PAH emission, and the $12.2\,\mu$m continuum.  
The shaded areas
  indicate the extraction apertures for 1D spectra 
(Figs.~1 and 2). The nuclei are 
located at the continuum peak.}
\end{figure*}

The presence of the 9.7\,$\mu$m silicate absorption feature in all of the
nuclear spectra is clear,  but its depth varies significantly from galaxy to
galaxy (see Fig.~\ref{f:fig1}). This feature is commonly seen in absorption in
Sy 2s  and in emission in some type 1 AGNs (Roche et al. 1991; 
Clavel et al. 2000; Shi et al. 2006). Shi et al. (2006) have found for Sys 
a relation between the depth of the silicate feature measured from {\it
  Spitzer}/IRS spectra  and
the hard X-ray column density. 
All the galaxies shown in Fig.~1, except IC~4518W for which there are 
no X-ray data, are Compton-thick so they should show similar depths. 
We note that although 
the limited T-ReCS spectral range  
makes it difficult to reliably estimate the continuum to measure the depth of
this feature, the observed range of depths seems to be real.

We have also marked in Fig.~1 the 8.6 and $11.3\,\mu$m PAH 
features. The Sy nuclei of NGC~5135 and IC~4518W do not show
PAH emission  as found for other
Sy nuclei  with high spatial resolution mid-IR spectroscopy 
(e.g., Circinus Roche et al. 2006 and NGC~1068 Mason et al. 2006). 
This supports  the scenario that the \PAH\, carriers are
evaporated in the presence of a hard ionization field such as
an AGN. The only exception,  NGC~7130,  shows extended  near
and mid-IR emission (Alonso-Herrero et al. 2006b), consistent with the
presence nuclear PAH due to SF.

The relatively high ionization potential (34.8\,eV) of the 
\SIV10.5\,$\mu$m  emission line 
may be interpreted as an AGN signature, but it has
also been detected in very young starbursts (Roche et
al. 1991; Verma et al. 2003) 
where the ionization is dominated by young hot stars. 
The line is detected in the three LIRGs (Fig.~1). 
The spatial distributions show that \SIV\, 
is clearly present in off-nuclear regions associated with the
AGN with an asymmetric distribution, especially in the case of IC~4518W 
(Fig.~3), as also found for Circinus (Roche et al. 2006).

\section{Spatially resolved mid-IR spectroscopy of NGC~5135}
\label{sec:ext}

In NGC~5135 the T-ReCS slit was placed such (Fig.~2) 
we could observe the Sy2 and  a bright H\,{\sc ii} region (with mid-IR and
Pa$\alpha$ emission) located $\sim 2.5\arcsec \sim 600\,$pc away, 
as well as the
region with diffuse emission between them. 
The {\it Spitzer}/IRS spectrum of the central $4\arcsec
 \times 4\arcsec$ ($\sim 1\,{\rm kpc} \times 1\,{\rm kpc}$) 
includes emission from  the Sy nucleus and a number of H\,{\sc ii} regions,
as well as  diffuse emission.
The \SIV\, emission line arises only from the nuclear regions (Figs.~2 and 3),
and it is not detected in 
the H\,{\sc ii} region, whereas the lower ionization  
[Ne\,{\sc ii}]$12.8\,\mu$m line
is detected in both (Fig.~2). There is $11.3\,\mu$m PAH feature 
in the H\,{\sc ii} region, and it is clearly extended 
in the diffuse region between 
the nucleus and the H\,{\sc ii} region. The spectrum of the H\,{\sc ii}
region shows  PAH feature emission, and [Ne\,{\sc ii}] emission.
In contrast the diffuse region
show faint continuum emission and no [Ne\,{\sc ii}] line,  and 
strong PAH emission with a large equivalent width (Fig.~2). 

Summarizing, spatially resolved mid-IR T-ReCS spectroscopy has allowed 
us to separate 
out different emission mechanisms such as nuclear emission (SF
and/or AGN), \HII\ regions, and regions of diffuse emission not associated
with strong ionizing sources. 
This kind of studies will be ideally suited for the GTC/CanariCam system.

Support was provided by 
the Spanish PNE (ESP2005-01480) and the NSF (0206617).
Based on observations obtained at the Gemini Observatory, which is operated by 
AURA, Inc., under a cooperative agreement
with the NSF on behalf of the Gemini partnership:  NSF (USA),  
PPARC (UK), NRC (Canada), CONICYT (Chile), ARC
(Australia), CNPq (Brazil) and CONICET (Argentina).


\begin{thebibliography}

\bibitem{} Alonso-Herrero, A., et al. 2006a, ApJ, 650, 835
\bibitem{} Alonso-Herrero, A., et al. 2006b, ApJL, in press, 
astro-ph/0610394
\bibitem{}Armus, L., et al. 2004, ApJS, 154, 178 
\bibitem{}Clavel, J. et al. 2000, A\&A, 357, 839
\bibitem{}Genzel, R., et al. 1998, ApJ, 498, 579
\bibitem{} Mason, R. E., et al. 2006, ApJ, 640, 612
\bibitem{} P\'erez-Gonz\'alez, P., et al. 2005, ApJ, 630, 82
\bibitem{} Roche, P. F., et al.  2001, MNRAS, 248, 606
\bibitem{} Roche, P. F., et al.  2006, MNRAS, 367, 1689
\bibitem{} Sanders, D. B. \& Mirabel, I. F. 1996, ARA\&A, 34, 749
\bibitem{} Shi, Y., et al. 2006, ApJ, in press, astro-ph/0608645
\bibitem{}Soifer, B. T., et al. 2002, AJ, 124, 2980
\bibitem{} Telesco, C. M., et al. 1998, Proc. SPIE, 3354, 534
\bibitem{}Verma, A., et al. 2003, A\&A, 403, 829
\end{thebibliography}
\end{document}